\documentclass[twocolumn,appendixfloats,tighten]{aastex6}
\usepackage{newtxtext,newtxmath}
\usepackage[T1]{fontenc}
\usepackage{ae,aecompl}

\usepackage{amsmath}	
\usepackage{amssymb}	

\usepackage{ctable}
\usepackage{url}
\usepackage{xspace}
\usepackage[normalem]{ulem}
\usepackage{hyperref}
\usepackage[all]{hypcap}
\usepackage{graphicx}

\def\msun{{\rm\,M_\odot}}

\def\msun{{\rm\,M_\odot}} 
\def\lsun{{\rm\,L_\odot}}

\newcommand{\kms}{\, {\rm km\, s}^{-1}}

\newcommand{\be}{\begin{equation}}
\newcommand{\ee}{\end{equation}}

\newcommand{\rvir}{r_{\rm{vir}}}

\def\h2{${\rm\,H_2}$}

\usepackage{color}

\begin{document}

\title{Ultra-light Dark Matter is Incompatible with the Milky Way's Dwarf Satellites}  
\author{Mohammadtaher Safarzadeh\altaffilmark{1,2} \& David N. Spergel\altaffilmark{3,4}}

\altaffiltext{1}{Center for Astrophysics | Harvard \& Smithsonian, 60 Garden Street, Cambridge, MA, 02138, USA,
  \href{mailto:msafarzadeh@cfa.harvard.edu}{msafarzadeh@cfa.harvard.edu}}
\altaffiltext{2}{School of Earth and Space Exploration, Arizona State  University, AZ, USA}
\altaffiltext{3}{Center for Computational Astrophysics, Flatiron Institute, Simons Foundation, 162 Fifth Avenue, New York, NY 10010, USA}
\altaffiltext{4}{Department of Astrophysical Sciences, Princeton University, Princeton NJ 08544, USA}

\begin{abstract}
The density profiles of dwarf galaxies are a highly varied set. If the dark matter is an ultra-light particle such as axions, then simulations predict a distinctive  and unique profile. 
If the axion mass is large enough to fit the ultra-faint dwarf (UFD) satellites($m\gtrapprox 10^{-21}$ eV), then the models do not fit the density 
profile of Fornax and Sculptor and are ruled out by more than $3-\sigma$ confidence.  
If the axion mass is in the mass range that can fit mass profiles of Fornax and Sculptor dwarf spheroidals, 
then its extended profile implies enormous masses ($\approx10^{11}-10^{12}\msun$) for the UFDs. These large masses for the UFDS are ruled out by more than $3-\sigma$ confidence by dynamical friction arguments. 
The tension would increase further considering star formation histories and stellar masses of the UFDs. Unless future ultra-light dark matter (ULDM) simulations with baryonic feedback show a significant change in the density 
structure of the halos, the current data is incompatible with the ULDM scenario. Relaxing the slope constraint from classical dwarf galaxies would lead to excluding ULDM with mass less than $6\times10^{-22}$ eV.
\end{abstract}

\section{Introduction}

Despite the remarkable success of cold dark matter (CDM) cosmology in explaining the large scale structure of the universe, CDM has suffered from three main potentially related shortcomings \citep{Bullock:2017hs}: 
(i) CDM simulations predict cuspy halo profiles while observations point to more core-like centers \citep{Bullock:2001hp,Gentile:2004ba}; 
(ii) the predicted stellar velocity dispersions are larger than observed in Milky Way's satellites \citep{BoylanKolchin:2012id}; 
and (iii) the number of sub-halos predicted in the CDM simulations far exceeds the observed number of luminous  Milky Way satellites \citep{Klypin:1999ej,Moore:1999ja}.
The origin of this discrepancy has been argued to be either due to baryonic processes \citep{Governato:2012cw,DiCintio:2013em,Pontzen:2014hd,Pawlowski:2015fm,Onorbe:2015jy} or changes to the nature of dark matter. 

One possible alternative to CDM is Fuzzy Dark Matter or Wave Dark Matter where the dark matter particle is Bose-Einstein condensate scalar field with de Broglie wavelength 
about the same size as the dwarf galaxies' core \citep{Goodman:2000hi,Hu:2000ea,Schive:2014bm,Schive:2014bp,Schive:2016ea,Hui:2017fq}. 
The scalar field is then well described by the coupled Schr{\"o}dinger and Poisson equations \citep{Widrow:1993je} and DM haloes are the ground-state solution of the system.
In such DM model, the cores are developed due to the quantum pressure arising from the uncertainty principle of particles occupying the same ground state which counters gravity.

The comparison between the observed core profiles of the dwarf galaxies with the predicted core profiles based on fuzzy dark matter 
has been made in recent works \citep{Schive:2014bm,Marsh:2015iw,Calabrese:2016fx,GonzalezMorales:2017br} and 
different values for the mass of the ultra-light boson have been achieved. 
Fitting for the observed mass radial slope of Sculptor and Fornax, \citet{Marsh:2015iw} conclude the mass of the ultra-light boson should be 
$m<1.1\times10^{-22}$ eV, and \citet{GonzalezMorales:2017br} arrive at $m<0.4\times10^{-22}$ eV at 97.5 percent confidence. 
Similarly \citet{Schive:2014bm} arrives at $m\approx 0.8\times10^{-22}$ by analyzing Fornax dSph.  
On the other hand, by considering two ultra-faint dwarf (UFD) galaxies (Draco II and Triangulum II), \citet{Calabrese:2016fx} 
conclude $m\sim3.7-5.6 \times 10^{-22}$ eV which is not in agreement with the limits found based on the density profiles of Fornax and Sculptor. 


In this paper we put together the data on the half-mass radius of the dwarf spheroidals (dSph) and UFDs, and the measured slopes of Fornax and Sculptor from published works \citep{Wolf:2010df,Walker:2011eg,Martin:2016gh,Martin:2016kq}. We show how the ensemble of data on dwarf galaxies appear to be incompatible with the predicted core profiles of ultra-light boson DM. 

In \S\ref{sec:core_profiles} we briefly discuss the core-profiles in Wave (Fuzzy) dark matter cosmology.
In \S\ref{sec:results} we compare the data to the analytic estimates of the core profiles in ultra-light boson DM halos. In \S\ref{sec:df} we discuss the upper limits on the halo mass of the dSph
galaxies based on dynamical friction arguments. In \S\ref{sec:sfh} we discuss the upper halo mass limits of UFDs based on their star formation histories. We summarize our results in \S\ref{sec:summary} and present the caveats.

\section{Halo profiles in wave DM cosmology}\label{sec:core_profiles}

Cosmological simulations of light-dark matter \citep{Schive:2014bp} find that the density profile of the innermost central region of the halos at redshift $z=0$ follows
\be
\rho_s(r)=\frac{1.9~(10~m_{22})^{-2}r_c^{-4}}{[1+9.1\times10^{-2}(r/r_c)^2]^8} 10^9 {\rm M}_{\odot} {\rm kpc^{-3}} \,,
\ee
where $m_{22}\equiv m/10^{-22}{\rm eV}$ is the DM particle mass and $r_c$ is the radius at which the density drops to one-half its peak value for a halo at $z=0$. 
This relationship is accurate to 2\% in the range $0 < r < 3 r_c$.

The enclosed mass at a given radius $r$ is:
\be
M( < r) = \int_0^r 4 \pi \rho_s(r') r'^2 dr' \,. 
\ee
$M_c \equiv M(<r_c)$ gives approximately the central core mass. This definition of core mass, makes up about 25\% of the total soliton mass, and $M(<3~r_c)$ makes up about 95\% of the total soliton mass.
Core mass or radius and the total mass of the halo, $M_h$, hosting the galaxy are related \citep{Schive:2014bp}:
\be
M_c\approx \frac{1}{4} M_h^{1/3}(4.4\times10^7 m_{22}^{-3/2})^{2/3} \,,
\label{eq:mc}
\ee
\be
r_c \approx 1.6 m_{22}^{-1} \Big (\frac{M_h}{10^9 {\rm M}_\odot}\Big)^{-1/3}  {\rm kpc} \,.
\label{eq:rc}
\ee

\begin{figure}
{\includegraphics[width=\columnwidth]{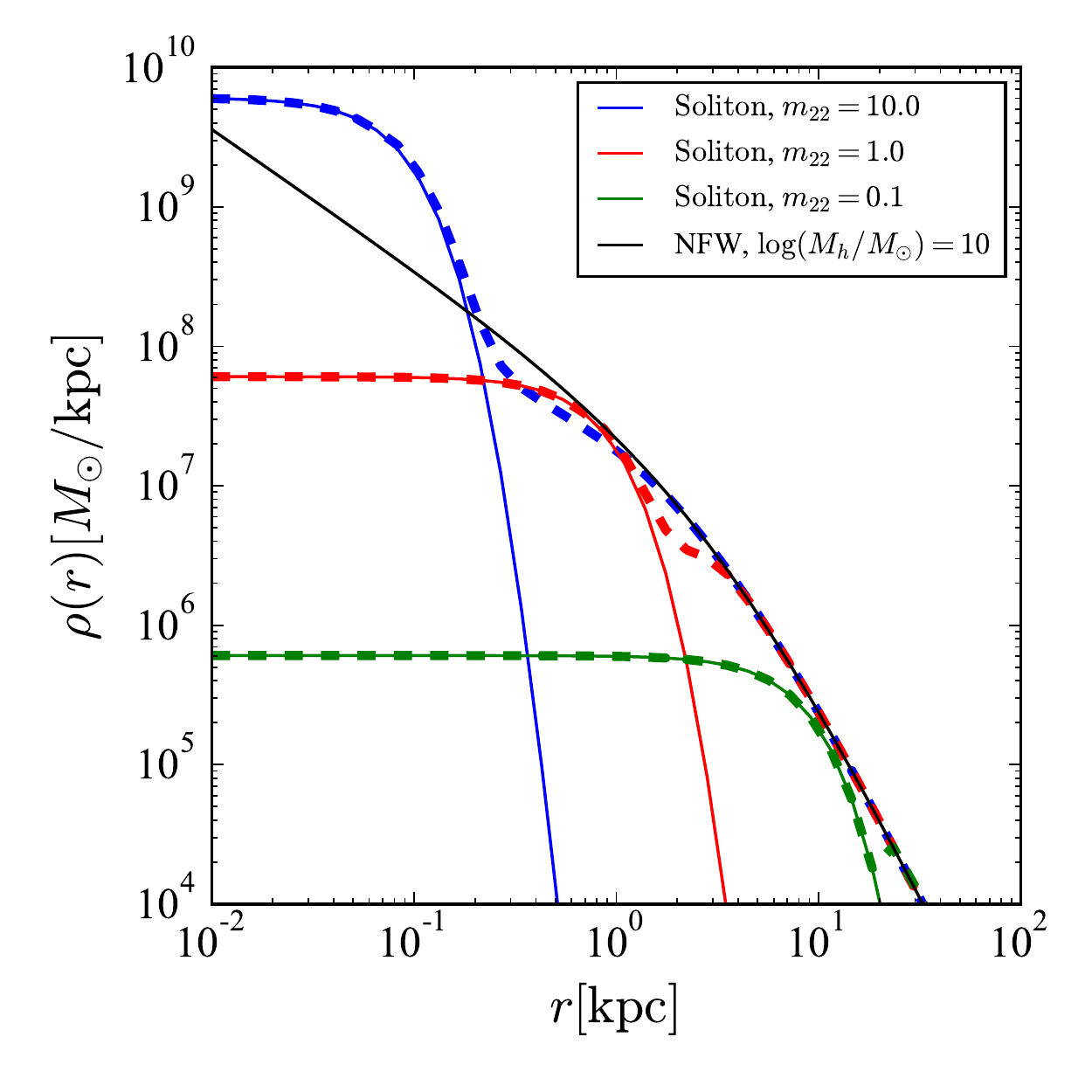}}
\caption{shows the modeled halo profiles of a $10^{10}\msun$ halo at $z=0$ for different values of $m_{22}$. 
Solid lines show the solitonic cores choice of $m_{22}$ (thin solid lines) and the thick dashed lines show the full halo profile that is a combination of the solitonic profile transitioning to an NFW
profile of mass $10^{10}\msun$ at around $r=3 r_c$.
}
\label{f:profiles}
\end{figure}

Beyond the core radius, the halo profiles resemble Navarro-Frenk-White \citep[NFW, ][]{Navarro:1997if} profiles \citep{Schive:2014bm}. 
We model each halo to have a central solitonic core profile which smoothly transitions to an NFW profile \citep{Mocz2018} around $r=3~r_c$.
We show the modeled profiles in Figure \ref{f:profiles}. Thin solid lines show the solitonic core profiles for different axion masses. The thin black line shows the NFW profile of a $10^{10}\msun$ halo at $z=0$.
The thick dashed lines show the full halo profile that is a combination of the solitonic profile transitioning to an NFW
profile of mass $10^{10}\msun$ around $r=3 r_c$.

\section{Comparison to Observational Data}\label{sec:results}

For a pressure supported system, one can use the Collisionless Boltzmann Equation (CBE) to related the six-dimensional (6D) phase-space distribution function, $f(\vec{r},\vec{v})$, of a tracer particle, 
to the underlying gravitational potential \citep{Binney:2008wd}. For nearby dwarfs we only have access to two spatial dimensions and one velocity dimension along the line of sight. 
dSph kinematic studies therefore rely on Jeans equations by integrating the CBE over velocity space:
\begin{equation}
  \frac{1}{\nu}\frac{d}{dr}(\nu \bar{v_r^2})+\frac{2}{r}(\bar{v_r^2}-\bar{v_{\theta}^2})=-\frac{GM(r)}{r^2},
  \label{eq:jeans}
\end{equation}
where $\nu(r)$ is the stellar density profile, and $\bar{v^2_r}$ and $\bar{v^2_{\theta}}$ are components of the velocity dispersion in radial and tangential directions, respectively.
The velocity anisotropy quantified by the ratio $\beta_{\mathrm{ani}}(r)\equiv 1-\bar{v^2_{\theta}}(r)/\bar{v^2_r}(r)$ is unconstrained by data.
Different anisotropic profiles can fit the projected velocity dispersion profile observed for the Fornax dSph, however, 
despite the presence of the degeneracy between mass and anisotropy, the predicted enclosed mass within about the dSph half-light radius is the same
among the different Jeans models \citep{Walker:2011eg}.

We take the enclosed mass within half-mass radius of most of the UFDs and dSph systems from \citet{Wolf:2010df}, where the two are related to the observed line of sight velocity dispersion by,
\be
M_{1/2}\approx \frac{3 <\sigma_{\rm los}^2> r_{1/2}}{G}.
\ee
The brackets indicate a luminosity-weighted average and $r_{1/2}$ is the 3D deprojected half-light radius.
The data points for Draco II and Triangulum II are from \citet{Martin:2016gh} and \citet{Martin:2016kq}, respectively.

The measured slopes come from recent observations that some dSphs have more than one stellar population. Each population independently trace the underlying gravitational potential. 
\citet{Battaglia:2006iz,Battaglia:2011ev} report the detection of a two component stellar system for both dSphs such that 
a relatively metal-rich subcomponent is more centrally concentrated with small velocity dispersion and a separate metal-poorer, kinematically hotter, more extended subcomponent. 
\citet{Walker:2011eg} measure the half-light radii and velocity dispersions of both subcomponents in Fornax and Sculptor, and effectively resolve two discrete points in a mass profile dominated by dark matter. 
\citet{Walker:2011eg} report the measured slope of the mass profiles defined as:
\be
\Gamma\equiv\frac{\Delta~{\rm log}~M}{\Delta~{\rm log}~r} = \frac{{\rm log} [M(r_{h,2})/M(r_{h,1})]}{{\rm log}[r_{h,2}/r_{h,1}]}
\ee
$r_h=3/4~r_{1/2}$ is the projected half-light radius.
The measured slopes for Fornax and Sculptor dSphs are $\Gamma=2.61^{+0.43}_{-0.37}$, and $2.95^{+0.51}_{-0.39}$ respectively \citep{Walker:2011eg} and they rule out cuspy NFW profiles ($d~{\rm log}M/d~{\rm log}r<2$ at all radii) with a significance  $>$96\% and  $>$99\%, respectively. 

\begin{figure*}
{\includegraphics[width=0.32\linewidth]{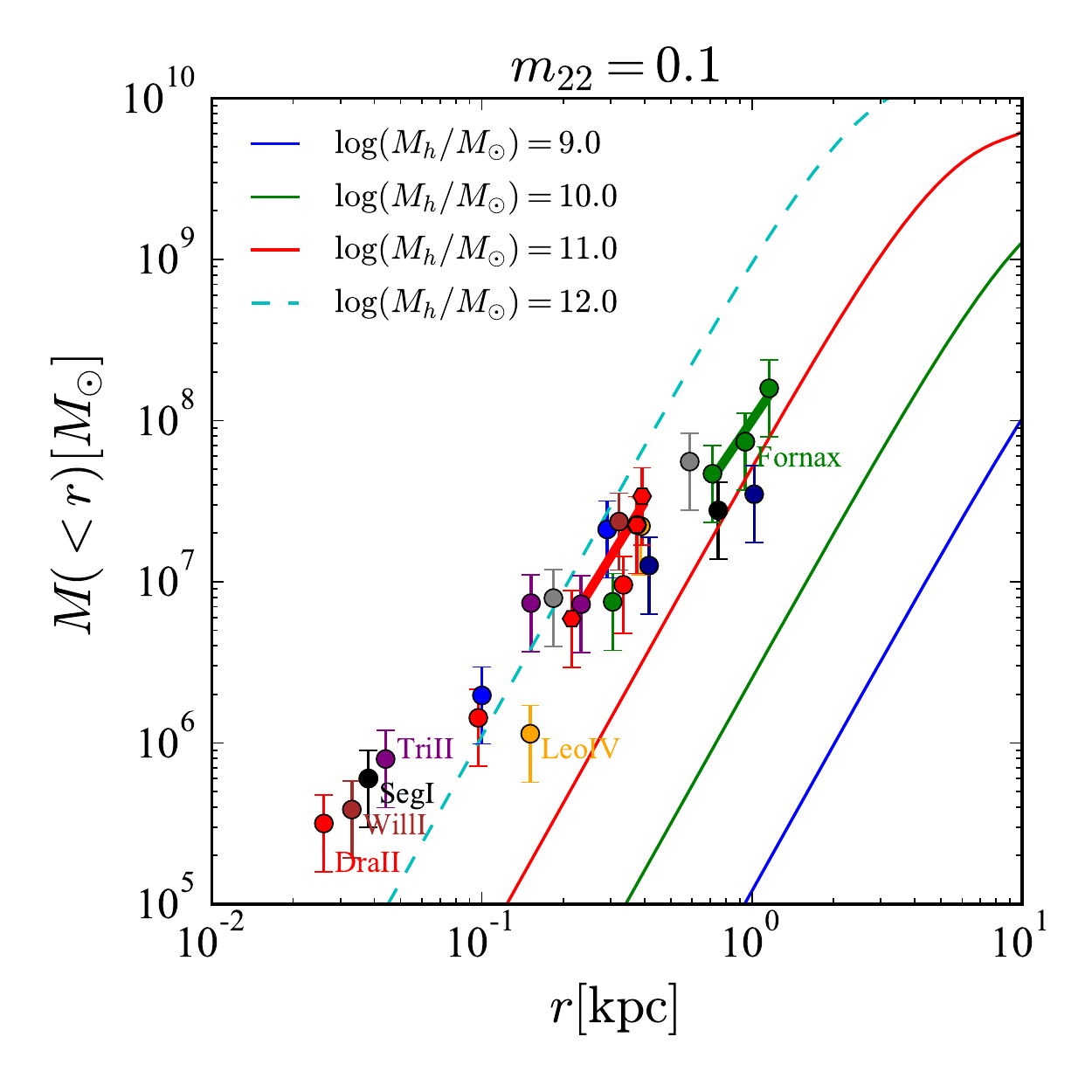}}
{\includegraphics[width=0.32\linewidth]{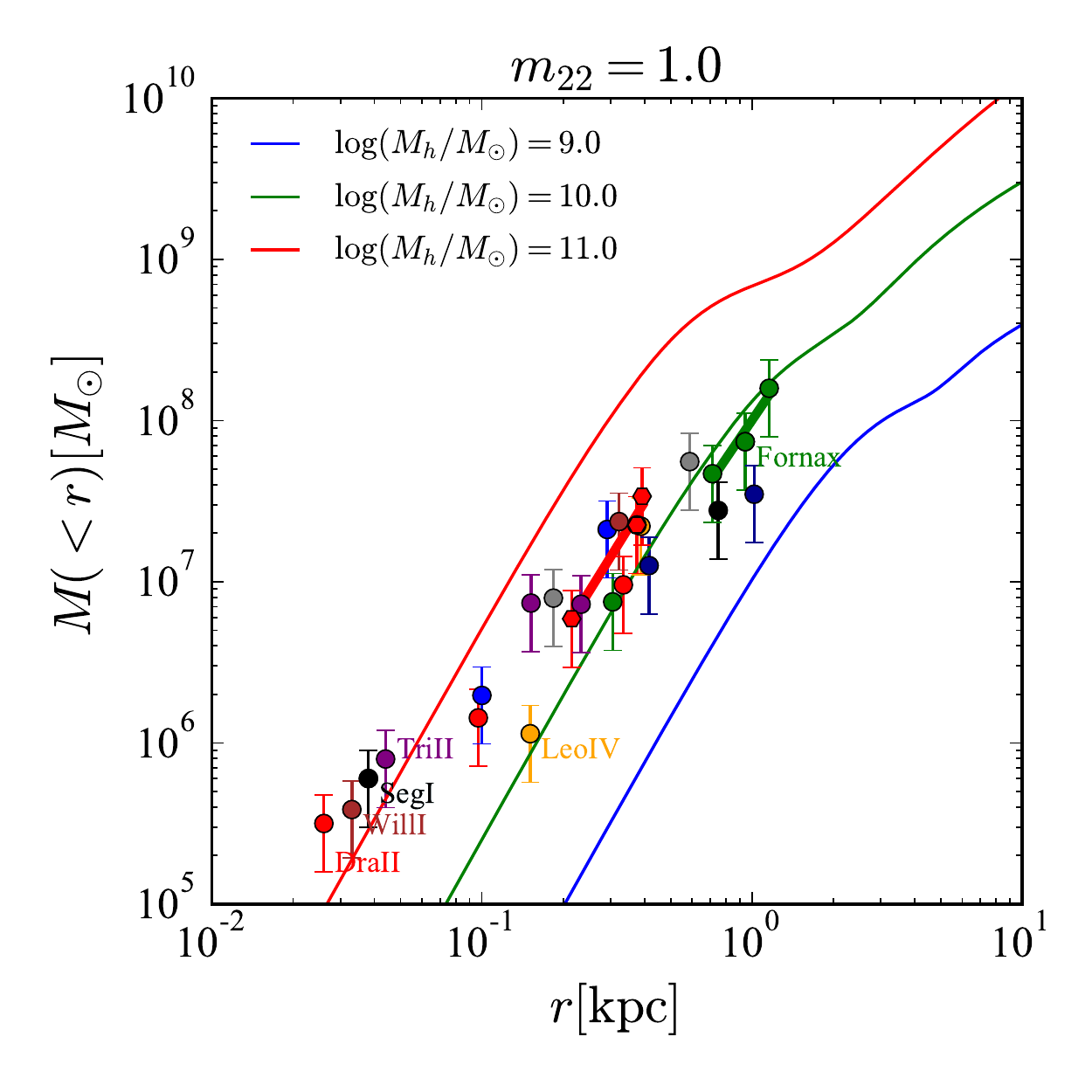}}
{\includegraphics[width=0.32\linewidth]{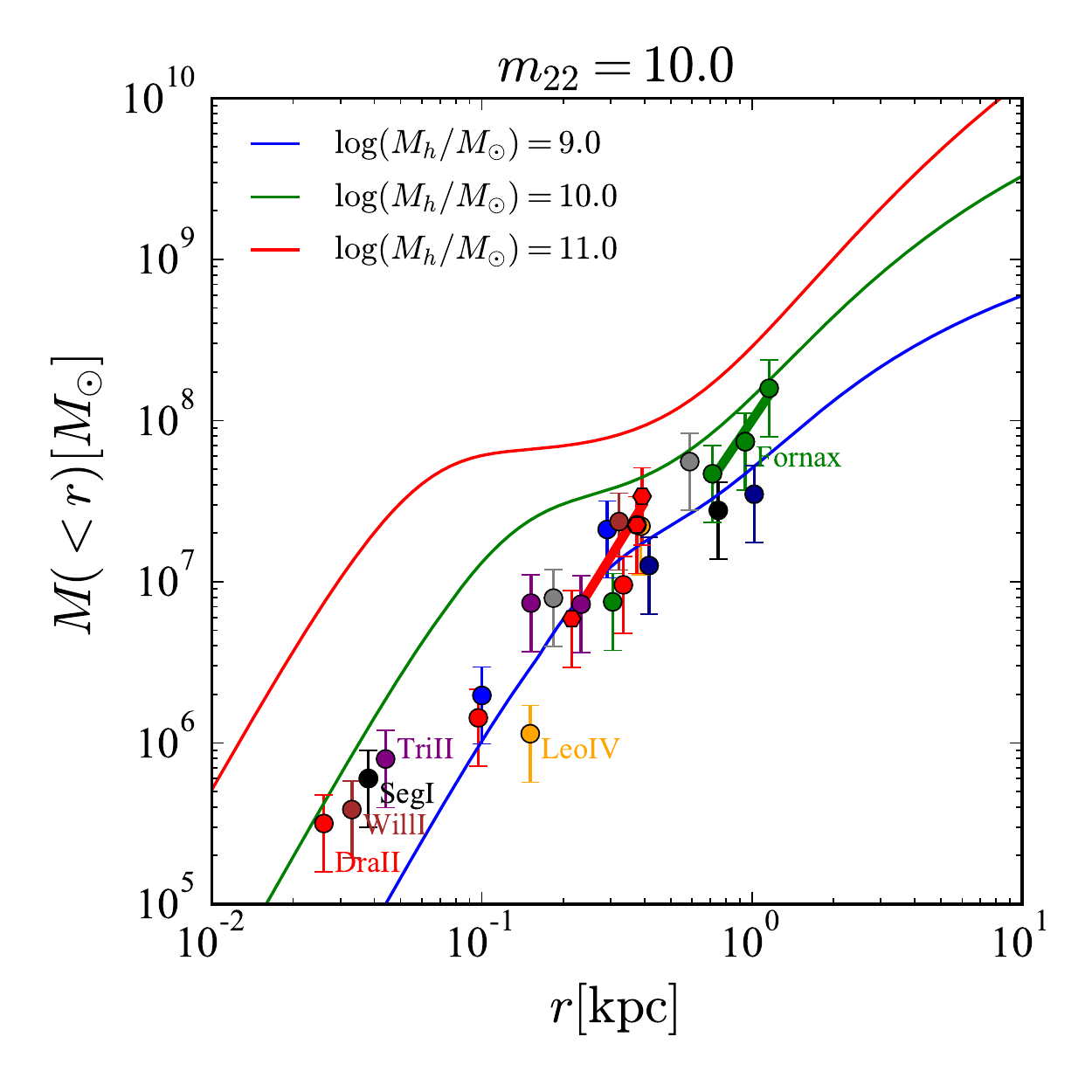}}
\caption{Comparing the parametrized mass profile of ultra-light dark matter against observations of the half-mass radius of dSphs and UFDs. 
In each panel we show the mass profiles corresponding to different \emph{total} halo masses as indicated in the legends. 
Left, middle, and right panels show the profiles corresponding to $m_{22}$=0.1, 1, and 10 respectively. 
The individual data points for the systems are collected from \citet{Wolf:2010df,Martin:2016gh,Martin:2016kq}, and the slopes of Fornax and Sculptor which are shown with green and red lines are from \citet{Walker:2011eg}. 
The errorbars are all inflated to be 0.1 dex. With $m_{22}$ less that 1, 
the predicted halo mass of the dwarf galaxies is too high given their dynamical state in the galaxy, 
and higher $m_{22}$ does not agree with the inferred slopes of Sculptor and Fornax.}
\label{f:fig1}
\end{figure*}

Figure \ref{f:fig1} shows the parametrized halo profiles based on simulations \citep{Schive:2014bp} against observations of the half-mass radius of UFDs and dSphs. We show the observed measured 
mass profile slopes of Fornax and Sculptor with green and red lines respectively. 
In each panel, we show the mass profiles corresponding to different total halo masses from $10^9 \msun$ (blue lines), to $10^{11}\msun$ (red lines). 
Left, middle, and right panels show the profiles corresponding to $m_{22}$=0.1, 1, and 10 respectively. 
The profiles show a core region parametrized by Eq (\ref{eq:mc}) which smoothly transition to an NFW profile at $r=3~r_c$.

Similar to the results of \citet{Schive:2014bm,Marsh:2015iw,GonzalezMorales:2017br}, low mass axions ($m<10^{-22}$ eV) can explain the observed mass profile slopes in Sculptor and Fornax. 
However, at such low masses, the predicted halo masses of the UFDs (such as Segue I) is too high given their dynamical state as we will return to in the next section. 
On the other hand, similar to the results of \citet{Calabrese:2016fx}, high mass axions ($m>10^{-22}$) can explain the halo masses of the UFDs such as Draco II, Triangulum II, and Segue I, however, 
such high masses would predict halo profiles that do not agree with the observed mass profile slope of Fornax and Sculptor. Therefore, either one can explain the observed half-mass radius of the UFD type systems and 
increase the tension with the observed mass profile slopes in dSphs, or explain the mass profiles and increase the tension with the predicted halo masses of the UFDs.

\begin{figure}
{\includegraphics[width=\columnwidth]{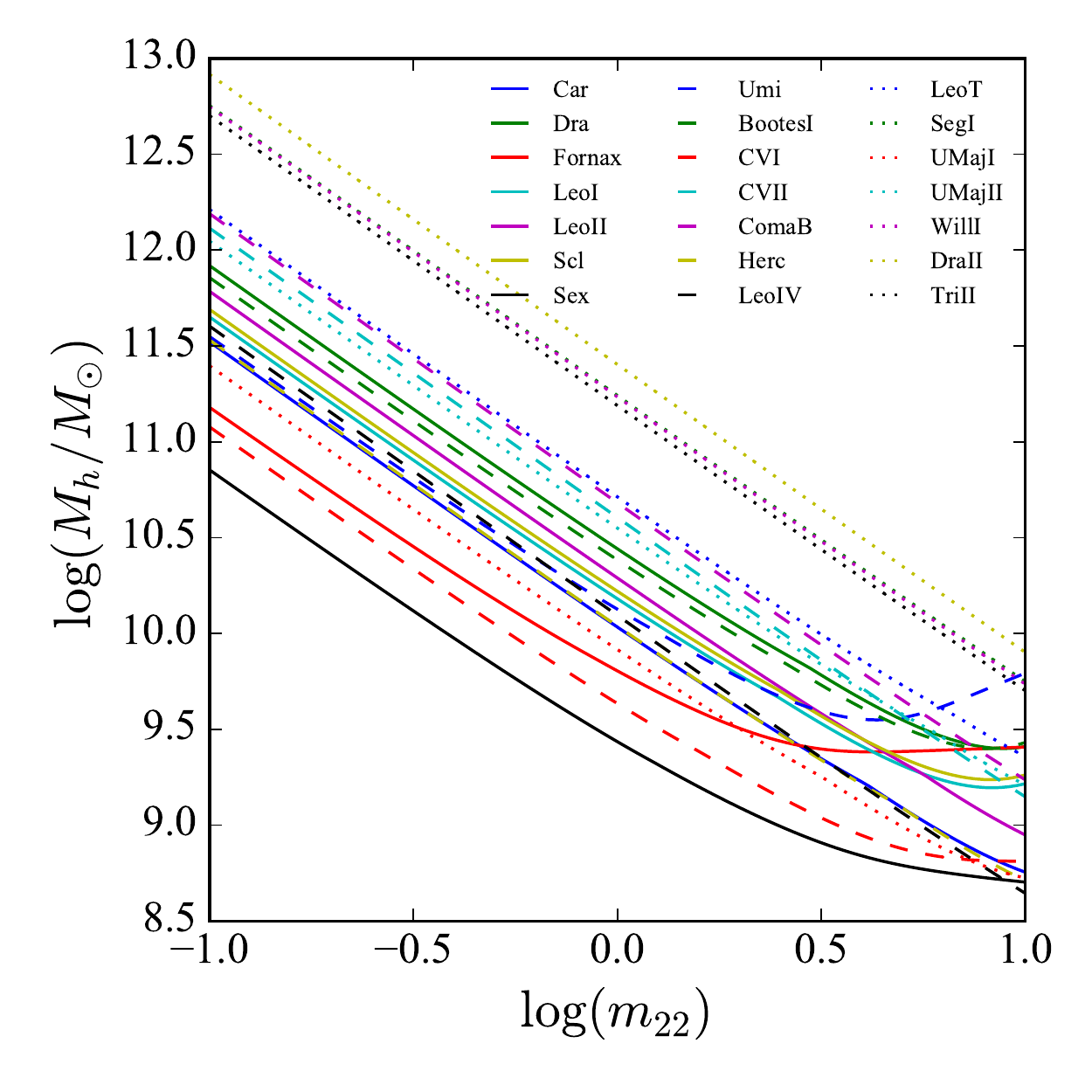}}
\caption{he estimated total halo mass of all the satellites with measured half-mass radius, as a function of $m_{22}$. 
Starting at $\log(m_{22})=-1$, increasing $m_{22}$ would predict lower halo mass for the satellites. However, the trend breaks at some values of $m_{22}$ and the estimated halo mass
increases again. The turning point indicates the start of NFW part of the profile to fit the observed data.}
\label{f:alt}
\end{figure}

Figure \ref{f:alt} shows the estimated halo mass of each of the dSph and UFDs as a function of axion mass. 
Starting at axion mass of $\log(m_{22})=-1$, increasing $m_{22}$ predicts lower halo mass for the satellites. However, the trend breaks at some values of $m_{22}$ and the estimated halo mass
increases again. The turning point indicates the start of NFW part of the profile to fit the observed data. 

Segue I and Willman I ($M_h\approx10^{10.5}\msun$) which is not only more than $3-\sigma$ away from the plausible halo mass for these satellites based on their dynamical friction timescale, 
but even more in tension with their halo masses inferred based on their stellar masses.  
Heavier axion masses predict a number density that falls below the observed lower limit and fails to fit the observed mass profile slopes of the Fornax and Sculptor dSphs. 



\section{Dynamical Friction Constraint}\label{sec:df}

The analysis of the orbits of the UFDs from the recent {\it Gaia} data release shows that other than Tucana III, all other UFDs have pericenters more than 20 kpc \citep{Simon:2018hp}. 
Therefore, since these UFDs lie far outside a potential core at the center of the MW, the dynamical friction timescale could be approximated assuming the satellites are in orbit in a host with a circular velocity of $v_c\approx200~\kms$.
The dynamical friction is computed as:
\be
\tau_{\rm fric}=\frac{10^{10} \rm yr}{\rm ln~\Lambda} \Big(\frac{r}{60 \rm~kpc}\Big)^2 \Big(\frac{v_c}{220 \kms}\Big) \Big(\frac{2\times10^{10}\msun}{M}\Big).
\ee  
$\Lambda=2 v_c/\sigma_s$, where $\sigma_s$ is the velocity dispersion of the satellite. 
By analyzing cosmological zoom-in simulations of 48 MW/ M31-like halos \citet{Wetzel:2015bs} conclude that a typical infall time for satellites of MW and M31 
is between 5-8 Gyr ago. We can arrive at an upper limit on the mass of the halo mass of the satellites requiring that $\tau_{\rm fric}$ be longer than infall time. 
By adopting a conservative approach and setting $r=\rvir$ of the host halo ($\approx200~\rm kpc$), and adopting the velocity dispersions from the recent compilation of \citet{Fattahi:2018cf}, 
we arrive at a firm upper limit of $\approx10^{11}\msun$.

\begin{figure}
{\includegraphics[width=\columnwidth]{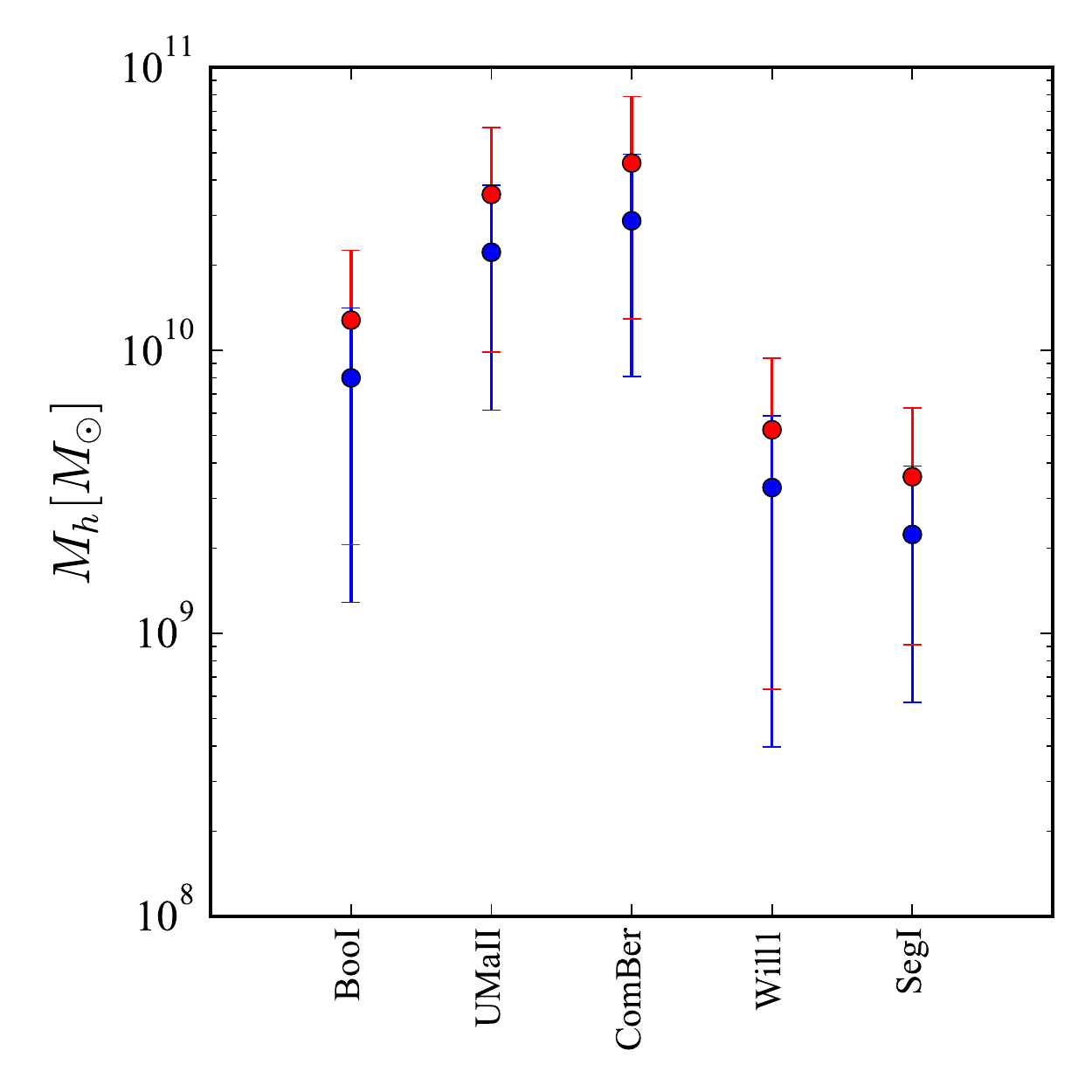}}
\caption{The upper limits on the halo mass of the UFDs with defined peri-center and apo-center distances from {\it Gaia} release \citep{Simon:2018hp}. 
 The red (blue) error bars show the results assuming the infall time for the satellites is 8 (5) Gyr ago. 
 The effective distance of the satellites are set to be their semi-major axis, as opposed to the virial radius of the host.}
\label{f:fig2}
\end{figure}

\citet{Simon:2018hp} measured the pericenter and apocenter of some of the UFDs from the recent {\it Gaia} data release. 
We compute the $\tau_{\rm fric}$ based on the orbit of these UFDs and we show the results in Figure \ref{f:fig2}. The arrived upper limits are more stringent as the orbits of these satellites show highly eccentric orbits
with an effectively shorter semi-major axis. Such upper limits for Segue I and Willman I rule out axion masses with $m_{22}<10$ with high confidence. 

\section{Star Formation History Constraint}\label{sec:sfh}

Another constraint on the axion mass of the halos hosting UFDs comes from their observed SFH. 
UFDs \citep{Brown:2012jo,Frebel:2012ja,Vargas:2013ei} are dark matter dominated galaxies \citep{Simon:2007ee} with total luminosities of $L_{\star}\approx10^3-10^5\lsun$. 
UFDs have very old stellar populations \citep[$>12\,$Gyr][]{Brown:2014jn,Weisz:2014cp} implying that they formed most, if not all, of their stars prior to reionization \citep[e.g.][]{Bullock:2000bn,Bovill:2011bk}. 

By tracking N-body simulations capable of resolving UFD host halos, based on four various abundance matching techniques, \citet{Safarzadeh:2018fi} showed the UFD host halos at $z=0$, would have 
a maximum plausible halo of mass of $10^8$ ($10^{9})\msun$ if the formation redshifts of these halos are considered to be $z=12~(6)$. 
These halos, will experience tidal stripping and therefore their masses today should be at most the same, and likely below their halo mass at the time of their formation. 


\section{Summary and Discussion}\label{sec:summary}

\begin{figure}
{\includegraphics[width=\columnwidth]{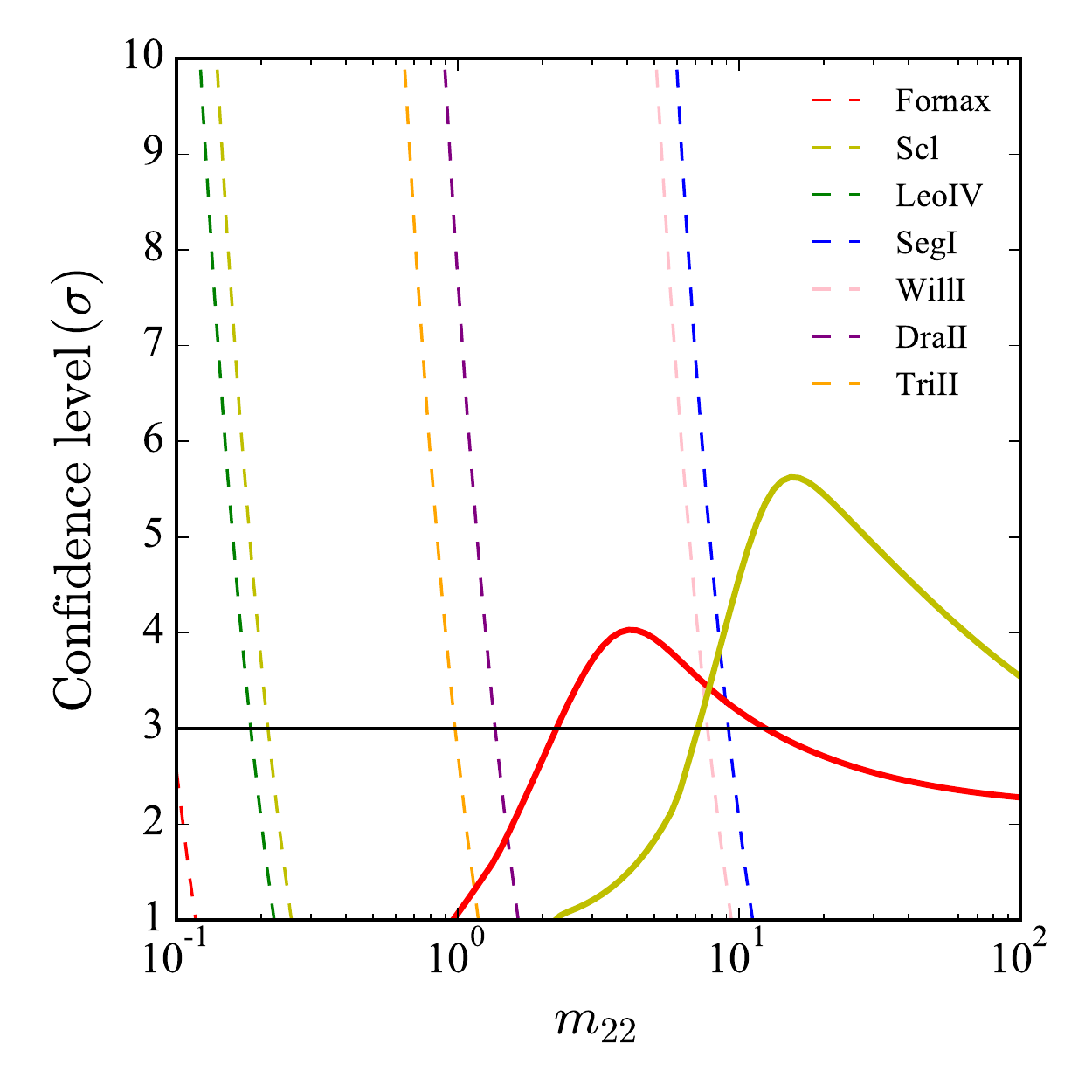}}
\caption{The confidence level by which a given axion mass is ruled out by any individual satellite. The thin lines are based on the predicted halo masses and their associated upper limits from dynamical friction timescales. 
The two thick green and red lines are based on the measured slopes of Fornax and Sculptor respectively. 
Based on Segue I and Willman I alone (the two right most dashed blue and pink lines) $m_{22}<8$ is ruled out with more than $3-\sigma$ confidence.
The measured Fornax and Sculptor slopes rule out $2<m_{22}<10$, and $m_{22}>10$ with more than $3-\sigma$ respectively.}
\label{f:summary}
\end{figure}

We summarize our results in Figure \ref{f:summary} where we show the confidence level by which a given axion mass is ruled out by any individual satellite. 
The horizontal black line shows the $3-\sigma$ limit. 
The thin lines each show $\sigma$ defined as $\sigma=(M_{h}(m_{22})-M_{\rm dyn})/\sigma_{\rm dyn}$ as a function of $m_{22}$, where $M_{\rm dyn}$ is the upper limit achieved when considering dynamical friction timescale of each satellite. 
The errors on the dynamical friction upper limits are assumed to be $\sigma_{\rm dyn}=0.2 M_{\rm dyn}$ to be on the conservative side. For satellites with available {\it Gaia} data such as Segue I and Willman I, the upper limits are much less
than $10^{11}\msun$.
The two thick green and red lines are based on the measured slopes of Fornax and Sculptor respectively defined as $\sigma=(\Gamma(m_{22})-\Gamma_{\rm WP})/\sigma_{\rm WP}$ where $\Gamma_{\rm WP}$ and $\sigma_{\rm WP}$ are
the slopes and the associated 1-$\sigma$ error from \citet{Walker:2011eg}.

Based on Segue I and Willman I alone (the two right most dotted pink and green lines) $m_{22}<8$ is ruled out with more than $3-\sigma$ confidence.
The measured Fornax and Sculptor slopes rule out $2<m_{22}<10$, and $m_{22}>10$ with more than $3-\sigma$ respectively.

The best fit mass value for ultra-light axions is $\sim 1 \times 10^{-21}$ eV.  
Even this value appears to imply too large a mass for the UFDs and seem to be a poor fit to Fornax and Sculptor. If future simulations suggest a steeper profile than 
\citet{Schive:2014bm}, then the conflict with Fornax and Sculptor observations could be alleviated. 
It would be helpful to reexamine this mass range to confirm our conclusions about the viability of the ultra-light dark matter models.

\citet{2018PhRvD..97f3507D} studied the tidal disruption of the sub halos ins fuzzy dark matter scenario. For a sub halo solitonic core to survive $N$ orbits the following condition should be satisfied:
\begin{align}
    M_c &>  5.82\times10^8 \left[\mu_{\rm min}(N_{\rm sur})\right]^{1/4} m_{22}^{-3/2} \left(\frac{D}{\rm kpc}\right)^{-3/4}\nonumber\\
        &\left(\frac{M_{\rm host}}{10^{12} M_{\odot}}\right)^{1/4} M_{\odot}\,.
\label{eq:M_c_min}
\end{align}

\begin{figure}
{\includegraphics[width=\columnwidth]{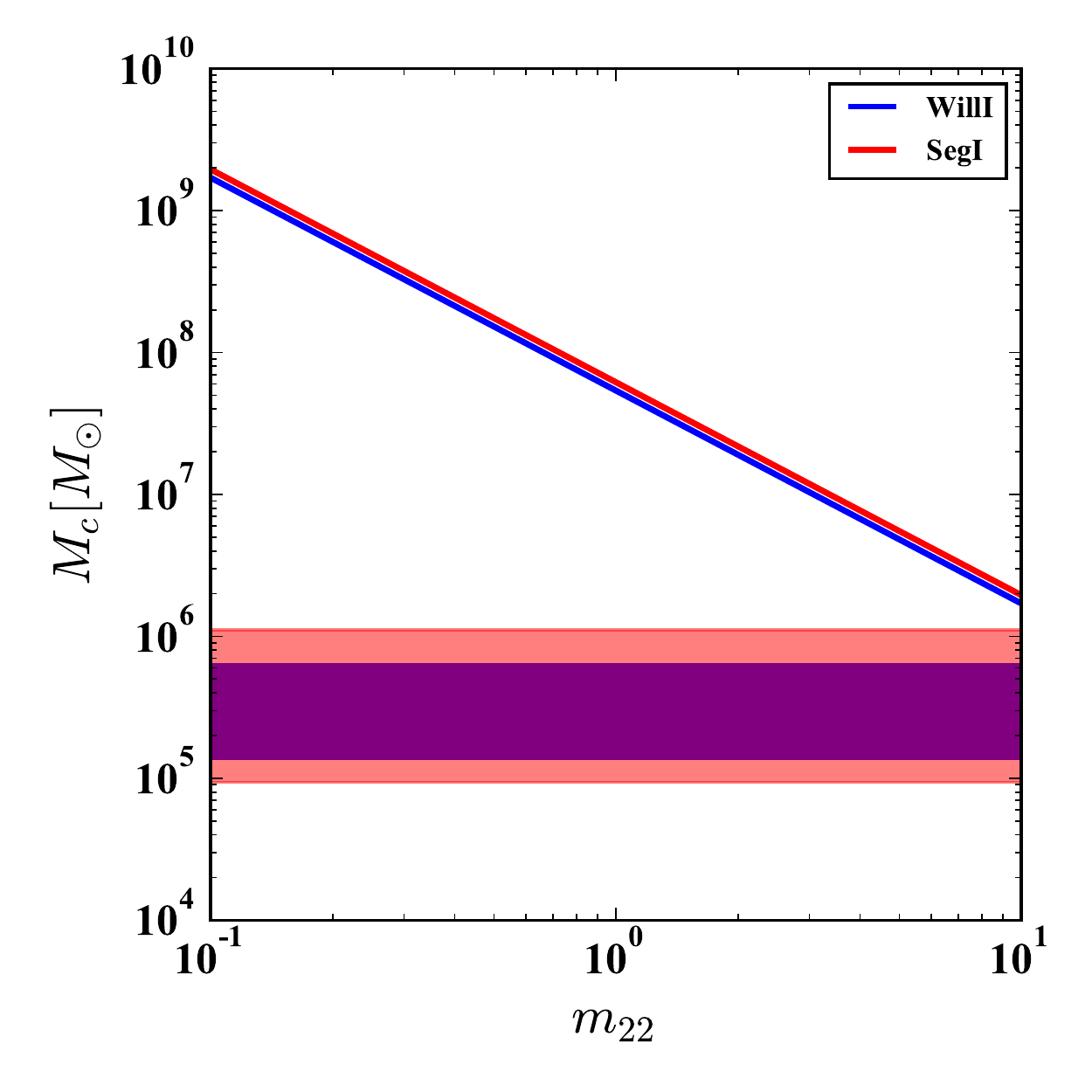}}
\caption{minimum core mass needed to survive one orbit as a function of $m_{22}$ for Seg I and Will I UFDs. The bands show the half-light mass of these two satellites.
$m_{22}<10$ is required for both these satellites to survive at least one orbit given their orbital parameters.}
\label{f:tf_fuzzy}
\end{figure}

Requiring that both Seg I and Will I survive at least one orbit, we set $\mu_{\rm min}=7.8$ which has been derived for the most conservative case (our result is not sensitive to this parameter). 
The two lines in Figure \ref{f:tf_fuzzy} show what solitonic core mass these two satellites should have in order to survive one orbit as a function of $m_{22}$.
The bands show the $M_{1/2}$ (half-light mass) of these two satellites. The results indicate that we need $m_{22}<10$ for these satellites to survive, which is consistent with the results we have obtained through dynamical friction argument.

We note that the core-halo mass relation in ultra-light dark matter (ULDM) simulations has a strong dependency, and since results of \citet{Schive:2014bp} 
suggest that there might be trends as a function of redshift for the adopted analytic relation that is not formulated yet, 
more simulation at low mass end at z=0 would be required to determine the validity of such scaling relations at low masses in the local universe.

Preliminary simulations of stars in ULDM simulations suggest lower core masses \citep{Chan2018}, and other lines of studies 
have suggested the importance of baryonic feedback effect \citep{Deng2018,Robles2019} in halos more massive than the UFD host halos. 
Future self-consistent simulations of baryonic feedback in ULDM cosmology 
would be the next step forward in determining the status of the simulations concerning the observations of the UFDs. 
However, in the case of halos hosting UFDs, given the small stellar mass of the UFDs, we anticipate the baryonic feedback to play a negligible role.

Two important caveats needs to be addressed in our work before a firm conclusion could be arrived at: (i) the binary fraction in the UFDs, and (ii) relaxing the classical dwarf galaxies' slope constraint.
We address these points below:

\subsection{Binary fraction in UFDs}
\begin{figure}
{\includegraphics[width=\columnwidth]{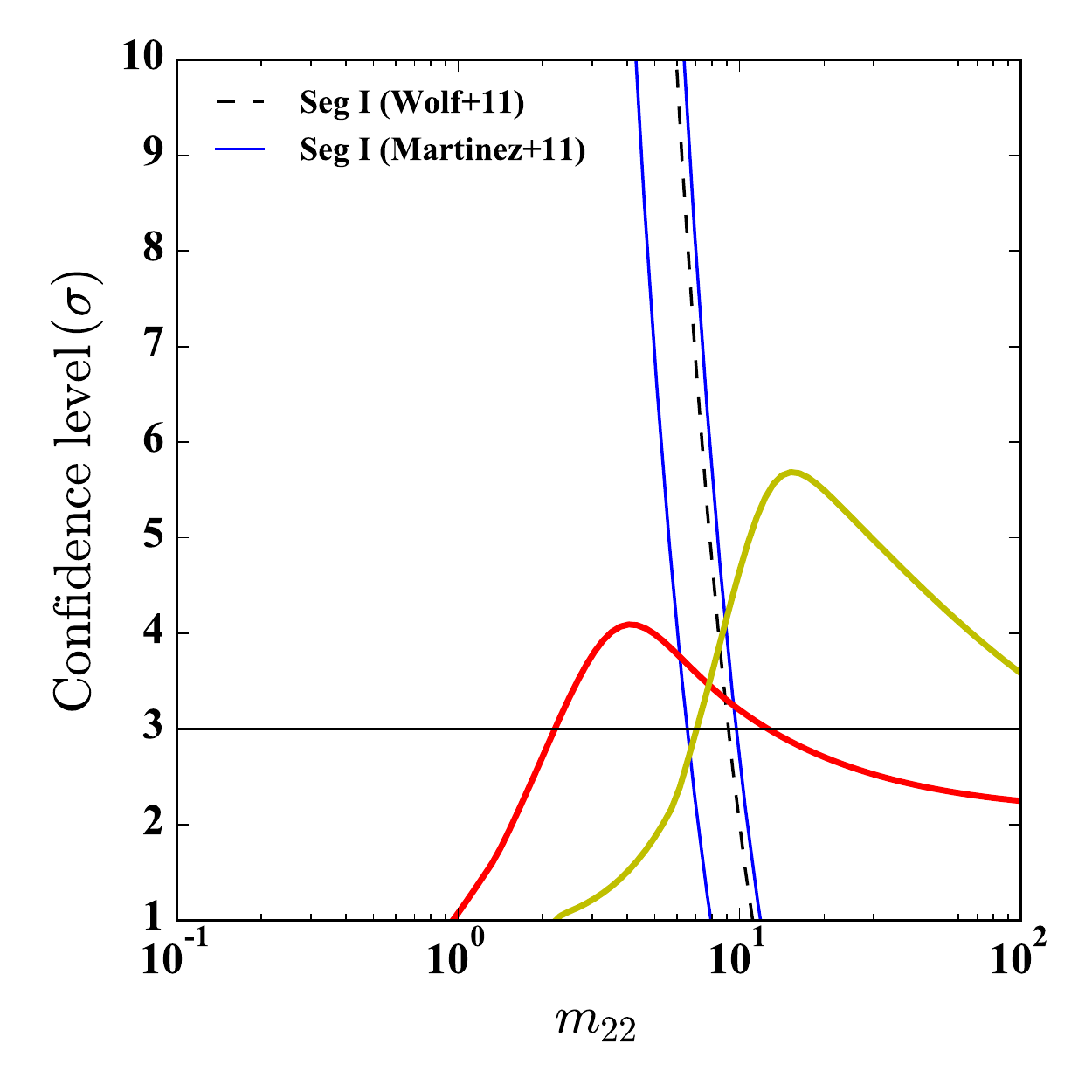}}
\caption{The same as in Figure \ref{f:summary} but focusing on Seg I. 
The exclusion boundary of Seg I given different assumptions regarding the contribution of the binary stars to its inferred intrinsic velocity dispersion. The black dashed line is based on the inferred half mass and radius reported 
in \citet{Wolf:2010df}. The blue lines delineate the boundary of exclusion region based on the work by  \citet{Martinez:2011ji} which corrects for the potential contribution of binary stars to the inferred mass and half light radius of Seg I.}
\label{f:segI}
\end{figure}

Presence of binary stars in UFDs can contribute to the measured velocity dispersion of such systems and therefore constraints on binary fraction and correcting for such an effect is crucial. 
However, contribution of binary stars to the observed velocity dispersion of the UFDs is not fully studied. In a 14-parameter likelihood model, \citet{Martinez:2011ji} corrected for the binary fraction of the stars 
in Seg I and arrived at $\sigma=3.7~\kms$ as opposed to $4.3~\kms$ of \citet{Wolf:2010df}. This translates into a $M_{1/2}\approx5.8^{+8.2}_{-3.1}\times10^5\msun$. 
Since among all the UFDs that we have studied in this work, Seg I is the most constraining, in Figure \ref{f:segI} we show the impact of such uncertainty on the exclusion interval of ULDM mass based on Seg I alone.

The black dashed line in Figure \ref{f:segI} is based on the inferred half mass and radius reported 
in \citet{Wolf:2010df}. The blue lines delineate the boundary of exclusion region based on the work by  \citet{Martinez:2011ji} which corrects for the potential contribution of binary stars to the inferred mass and half light radius of Seg I.
Larger contribution of the binaries would result in a larger correction for the observed dispersion of the system which would translate to inferring the dynamical mass is less than what is observed. 
The left most blue line in Figure \ref{f:segI} indicates the boundary assuming the half-mass is $2.7\times10^5\msun$ and $r_{1/2}=36$ pc. This indicates that after correcting for contribution of the binaries to the observed 
velocity dispersion in Seg I, the ULDM with mass less than $m_{22}<6-10$ is excluded only based on Seg I alone.

\subsection{Removing the slope constraint}
The constant density core that is arrived in the analysis by \citet{Walker:2011eg} has been challenged in later studies. 
For example, by using dwarf galaxies in the APOSTLE (A Project Of Simulating The Local Environment) Lambda cold dark matter cosmological hydrodynamics simulations of analogues of the Local Group, 
\citet{2018MNRAS.474.1398G} conclude that while all the halos in the simulation box have NFW cusps, the \citet{Walker:2011eg} procedure often leads to a statistically significant detection of a core and 
the main cause is a violation of the assumption of spherical symmetry upon which the mass estimators are based. As a result, a wide range of slopes of the density profile could be inferred depending on the viewing angle.

Moreover, different approach to modeling the observations has lead to a contrary result. 
For example, modeling Sculptor data with distribution function of an equilibrium spherical system that is separable in energy and angular momentum would indicate the two metal poor and metal rich populations are in equilibrium within an NFW dark matter potential with structural parameters in the range expected in Lambda-CDM \citep{2017ApJ...838..123S}.
In a more recent work, \citet{2018ApJ...860...56S} show that velocity dispersion profiles of metal-rich and metal-poor stars provide good fits to the observed number count and velocity dispersion profiles of metal-rich and metal-poor stars both in cored and in cusped potentials. To be able to distinguish between the core and cusp will require velocity dispersion data with uncertainties well below 1 $\kms$ over a range of projected radii.

If we relax our assumption about the the slope constraint from \citet{Walker:2011eg} Seg I and Willman I provide the most stringent exclusion boundary for the ULDM mass and a limit of $m_{22}<6$ is arrived in the most conservative 
case based on Seg I alone.

The result of our work could be summarized as UFDs such as Segue I are much denser than Fornax. 
Any model where dark matter dynamics sets a universal core profiles cannot fit both systems. Baryonic physics must play
a significant role in shaping the profiles or the profiles must be sensitive to initial conditions.  
Because of their wave-like nature sets a characteristic scale, ultra-light dark matter models appear to be ``too predictive" and seem to be in conflict with the data. 


\acknowledgements 

We are thankful to the referee for their constructive comments. 
This work was supported by the National Science Foundation under grant AST14-07835 and by NASA under theory grant NNX15AK82G as well as a JTF grant. 
MTS is grateful to the Center for Computational Astrophysics for hospitality during the course of this work.

\end{document}